\documentclass[10pt,a4wide]{article}
\usepackage{newlfont}
\usepackage{amsmath}
\usepackage{graphics}
\usepackage{float}
\usepackage{graphicx}
\usepackage{epsfig}

\begin{document}
\title{Wave Function Based Characteristics of Hybrid Mesons}
\author{Nosheen Akbar\thanks{e mail: noshinakbar@yahoo.com}\quad,
Bilal Masud\thanks{e mail: bilalmasud.chep@pu.edu.pk}\quad, Saba Noor\thanks{e mail: sabanoor87@gmail.com}\\
\textit{Centre For High Energy Physics, University of the Punjab,
Lahore(54590), Pakistan.}}
\date{}
\maketitle

\begin{abstract}
We propose
some extensions of the quark
potential model to hybrids, fit them to the lattice data and use them for the purpose of calculating the
masses, root mean square radii and wave functions at the
origin of the conventional and hybrid charmonium mesons. We treat
the ground and excited gluonic field between a quark and an
antiquark as in the Born-Oppenheimer expansion, and use the shooting
method to numerically solve the required
Schr$\ddot{\textrm{o}}$dinger equation for the radial wave
functions; from these wave functions we calculate the mesonic properties.
For masses we also check through a Crank Nichelson discretization.
For hybrid charmonium mesons, we consider the exotic quantum number
states with $ J^{PC} = 0^{+ -}, 1^{- +}$ and $2^{+ -}$. We also
compare our results with the experimentally observed masses and
theoretically predicted results of the other models. Our results
have implications for scalar form factors, energy shifts, magnetic
polarizabilities, decay constants, decay widths and differential cross
sections of conventional and hybrid mesons.
\end{abstract}

%%% ----------------------------------------------------------------------
\maketitle
%%% ----------------------------------------------------------------------
\section*{I. Introduction}

\qquad A number of hadron properties are well described by the quark
model where mesons have quantum numbers $J=L\oplus S$, $P =
(-1)^{L+1}$ and $C = (-1)^{L+S}$, $L$ and $S$ being the quantum
numbers for the quark-antiquark orbital angular momentum and their net spin angular momentum
respectively.
The states with $J^{PC} = 0^{+-}, 1^{-+}, 2^{+-}$ (for
the lowest lying hybrids in the flux tube model) can not be formed
from a $q\overline{q}$ pair and hence are not allowed in the quark
model. These states are signals for exotic mesons (hybrids,
glueballs, etc). Quantum Chromodynamics (QCD), describing the
interaction between quarks and gluonic field, predicts the
existence of hybrid mesons containing the excited gluonic field.
Thus for understanding of QCD, we need to find experimentally
testable predictions of the theory for situations in which the
gluonic field between a quark and antiquark is in an excited state.
Thus hybrids are an important source of information related to
confining properties of QCD, and checking for existence of hybrid
mesons is very important objective of particle physics. Reviews of
the spectrum of excited gluonic states can be found in
ref.~\cite{McNeile02}. Recently, a resonance is
observed at COMPASS~\cite{M. Alek} with $J^{PC} = 1^{-+}$. Some
other groups like VES~\cite{Gouz}-\cite{Gouz2}, E852~\cite{Adams}-\cite{Adams4}, and the
Crystal Barrel collaboration~\cite{Baker} also observed these
states.

Using the Born-Oppenheimer approach,
the work of finding implications of QCD for a meson can be split into first using the
numerical lattice simulations of QCD to fit parameters of
a quark antiquark
potential and then using this potential to calculate its dynamical
implications. Even a numerically defined potential can be used in
this scheme, as in  refs.~\cite{Morningstar03}~\cite{Morningstar97}.
These works use some potentials to calculate hybrid masses and few radial probability graphs (but presents no $q\bar{q}$ wave function expressions or uses.).
What we add to this work is that now we suggest a number of analytical expressions for the excited state gluonic field potential between a
quark and antiquark and fit their parameters to the lattice data for the ground and
excited state gluonic field energy values available in
ref.~\cite{Morningstar03} for discrete quark antiquark separations. For each case, we report a dimensionless chisquare and other measures
directly telling how much fractional error each model has in fitting.
We use this variety of potentials to indicate suitable ones for different applications,
 for example for analytical calculations of expectation values in few-body wave functions where a numerical approach may have convergence problems. Our full list of chisquare and other error measures tells how much extra error is generated in preferring the flux tube model and other integrable forms over the ones that can be only numerically used but fit better. After using
  these potentials to find the quark-antiquark wave functions, we also calculate   a number of conventional and hybrid meson rms radii and square of radial wave functions at the origin.
We continue till writing some implications of these for scalar form factors, energy shifts, magnetic
polarizabilities, decay constants, decay widths and differential cross
sections of conventional and hybrid mesons. All this is addition to our own results for meson masses and radial probabilities
that are in agreement to those reported in ref.~\cite{Morningstar03} for the height of the single peak of the probability graph.

In the present paper, we apply our techniques to charmonium mesons.
An advantage of considering them is indicated in ref.~\cite{S.
Collin} as
\begin{quote}
 `` The best systems for a
hybrid search may be $c\overline{c}$ or $b\overline{b}$ where there
is a large gap between the lowest states and the $D\overline{D}$ and
$B\overline{B}$ threshod respectively.''
\end{quote}

To find the wave functions of conventional charmonium mesons, we use
the realistic columbic plus linear potential model to solve the
Schr$\ddot{\textrm{o}}$dinger equation numerically by using
 the corresponding quantum numbers of mesons. To study hybrids, we repeat
the numerical work with the models of the gluonic excitation energy mentioned in section III. From the numerically
found wave functions, we calculate the root mean square radii. These
radii can be used to find scalar form factors ~\cite{Ananthanarayan} for charmonium mesons, along with energy shifts~\cite{S. I. Kruglov}
and magnetic polarizabilities~\cite{S. I. Kruglov}. Thus we have
reported some predictions about these quantities for conventional
and hybrid charmonium mesons. We have also found the numerical values
of square of radial wave functions at the origin
 ($|R(0)|^2$), which can be
used to calculate the decay constants~\cite{Bhavin}, decay
rates~\cite{Bhavin}, and differential cross sections~\cite{diff} for
quarkonium states. The predictions about these quantities are also
reported for conventional and hybrid charmonium mesons.

In the section II below, we write the Hamiltonian for the
conventional mesons. Then we describe the shooting method-based numerical procedure to find the
solution of the radial
Schr$\ddot{\textrm{o}}$dinger equation for
conventional charmonium mesons. The expressions to find
masses, root mean square radii, and squares of radial wave functions at the origin
($|R(0)|^2$) of conventional charmonium mesons are also written in this section. In
section III, the Hamiltonian is written for hybrid mesons, and then
we redo all the numerical work as done in
section II for hybrids now.
The $\chi^2$ and other error measures
for different forms of the potential difference between
ground and excited state are also written in section III. Results
for the masses, root mean square radii and $|R(0)|^2$ of
conventional and hybrid mesons are reported in section IV for
systems composed of charm quarks and antiquarks. Based on these
results, we also include some results related to experimentally
measurable quantities.

\section*{II. Characteristics of Conventional charmonium mesons}
\subsection*{The Potential Model for Conventional Charmonium
Mesons}

\qquad In the potential models, the confining potential for
$Q\overline{Q}$ system in the  the ground state gluonic field is mostly used in the form of
\begin{equation}
\frac{-4 \alpha_{s}}{3 r}+ b\, r + c,\label{P21}
\end{equation}
with inter-quark distance $r$. Here, $- 4/3$ is due to the colour
factor, $\alpha_{s}$ is the quark-gluon coupling, $b$ is the string
tension and $c$ is a constant. In above equation, the first term is
due to one gluon exchange and the second term is the linear
confining potential~\cite{Perkin}. This potential form provides a
good fit to the lattice simulations of refs.~\cite{Bali,Bali1,Alexandrou}.
By including the Gaussian-smeared
hyperfine interaction~\cite{charmonia05} and orbital angular
momentum (or centrifugal) term, the potential of the
$Q\overline{Q}$ system for the ground state gluonic field have
following form
\begin{equation}
V(r)= \frac{-4 \alpha_{s}}{3 r}+ b\, r + \frac{32 \pi \alpha_{s}}{9 m^{2}_{c}}
(\frac{\sigma}{\sqrt{\pi}})^{3} e^{-\sigma^{2} r^{2}} S_{c}.S_{\overline{c}} + \frac{L (L + 1)}{2 \mu \, r^{2}},\label{P24}
\end{equation}
where $S_{c}.S_{\overline{c}}= \frac{S (S+1)}{2} - \frac{3}{4}$, $\mu$ is the reduced mass of the quark and antiquark, $m_{c}$ is the mass
of the charm quark, and $S$ is the total spin quantum number of the meson.
 For the $c\overline{c}$ mesons, the parameters
$\alpha_{s}$, $b$, $\sigma$, and $m_{c}$ are taken to be $0.5461$,
$0.1425 \textrm{GeV}^{2}$, $1.0946$ GeV and $1.4796$ GeV respectively
 as in ref.~\cite{charmonia05}.
The quantum numbers for the conventional
charmonium mesons we choose for our study are reported below in Table 5.

\subsection*{Wave Functions and Radii of Conventional Charmonium Mesons}

\qquad A conventional meson can be described by the wave function of the
bound quark-antiquark state which satisfies the
Schr$\ddot{\textrm{o}}$dinger equation with potential of
eq.\eqref{P24}. Radial Schr$\ddot{\textrm{o}}$dinger equation with
wave function $U(r) = r R(r)$ is written (in natural units)
as
\begin{equation}
\frac{d^2}{d r^2} U(r) + 2 \mu (E - V(r)) U(r) = 0.\label{P23}
\end{equation}
Here $R(r)$ is the radial wave function,  $r$ is the interquark distance, $E$ is the sum of kinetic and potential energy of quark-antiquark system,
and $V(r)$ and $\mu$ are defined above through eq.\eqref{P24}.

In quark-antiquark bound
state, the wave function must satisfy the boundary conditions $U(0) = 0$ and $U(\infty) = 0$.
For the numerical solution of the Schr$\ddot{\textrm{o}}$dinger equation
with the potential of eq.\eqref{P24}, we repeatedly generated energy E from -2 to 2 GeV in steps of 0.1 GeV. For each such trial initial energy, we used the Newton method~\cite{Newton}
to select, if any, the energy for which the numerical
solution of Schr$\ddot{\textrm{o}}$dinger equation became zero at
infinity.
To obtain these numerical solutions, we used the RK method~\cite{Vedamurthy} with using any arbitrary integer value of $U'(0)$. For different values of $U'(0)$, normalized solutions of the Schr$\ddot{\textrm{o}}$dinger equation, obtained by multiplying the solution
with the normalization constant ($\frac{1}{\sqrt{\int U^{2}(r) dr}}$), remain the same.
These energy eigenvalues plus constituent quark masses
are taken to be the $c\overline{c}$ mesons masses (in natural units). It
is found that our results for conventional charmonium mesons agree
with the Table 1 of ref.~\cite{charmonia05}. This supports the
reliability of our method. We also checked the consistency of our
method by 1) getting a $100\%$ overlap of our $H U$ and $E U$ and 2) by
calculating the masses of conventional mesons by the Crank Nichelson
Discritization and finding that masses obtained by both of the methods are identical.
\begin{figure}
\begin{center}
\epsfig{file=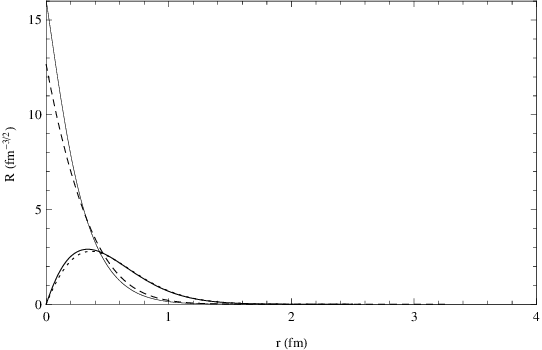,width=0.6\linewidth,clip=}\caption{The radial parts of $\eta$,$J/\psi$,$h_{c}$, and $\chi_{c}$ meson wave functions as functions of
$r$.
Thin solid line represents the $\eta$ wave function, dashed line represents the $J/\psi$ wave function, thick solid line represents the $h_{c}$ wave function and points represent the $\chi_{c}$ wave function.}
\end{center}
\end{figure}
The Fig.1 shows the dependence of $\eta$, $J/\psi$, $h_{c}$, and
$\chi_{c}$ normalized radial wave functions on the radial separation $r$. The
quantum numbers ($L$ and $S$) for these particles are given below
in Table 5. These graphs show that the radial wave functions of
$\eta$, $h_{c}$, $\chi_{c}$, and $J/\psi$ have the same properties
as that of hydrogen atom radial wave functions, i.e. they behave as
$r^{L}$ for small inter quark distances and decrease exponentially
at large inter quark distances. Thin solid and Dashed lines graphs
are for $L=0$, so these graphs are similar to $r^{0} exp(-r)$. Thick
solid and points are for $L=1$, so these graphs are similar to
$r^{1} exp(-r)$.
As $L$ increases, the wave function's
peak goes away from the origin. This means that centrifugal term has
more effects on wave function than that of the hyperfine term.
One possible reason is that we
are dealing with heavy quarks so the $1/m^2_{c}$
factor (shown in eq.\eqref{P24}) of the hyperfine term becomes very small.

The normalized wave functions are used in the further calculations for root mean square radii and square of radial wave
functions at origin. To find the root mean square radii of the $c\overline{c}$ mesons, we used
the following relation:
\begin{equation}
\sqrt{\langle r^{2}\rangle} = \sqrt{\int U^{\star} r^{2} U dr}.\label{P25}
\end{equation}
In ref.~\cite{M. Feyli}, for normalized wave function
\begin{equation}
U'(0) = R (0) = \sqrt{4 \pi}\, \psi(0) \label{mid}
\end{equation}
is used and we use this prescription.  Thus the derivative of $U(r)$ at $r = 0$ is calculated
 to find $|R(0)|^2$.
$|R(0)|^2$ is used in many applications of high energy
physics as mentioned in section I.

\section*{III. Characteristics of Hybrid Charmonium mesons}
\subsection*{The Potential Model for Hybrid Charmonium
Mesons}

\qquad The centrifugal factor for the hybrid mesons is written in
refs.~\cite{Juge99,Kuti97} as
\begin{equation}
\frac{L (L + 1) - 2 \Lambda^{2} + <J^{2}_{g}>}{2 \mu \, r^{2}} \label{P28}
\end{equation}
where $\Lambda $ is the projection of the total angular momentum
$J_{g}$ of the gluonic field. The states with $\Lambda = 0, 1, 2, 3,
,....$ are usually represented by the capital greek letters $\Sigma,
\Pi, \Delta, \Phi,...$ respectively. We are interested in finding
the masses and root mean square radii of the hybrid states $0^{+-},
1^{-+}$, and $2^{+-}$. These states can be generated from the
$\Pi_{u}$ potential. For the $\Pi_{u}$ potential, $<J^{2}_{g}> =
2$ and $\Lambda = 1$~\cite{Kuti97}.
Therefore $- 2 \Lambda^{2} + \langle J^{2}_{g} \rangle = 0$, so centrifugal factor for the hybrid
mesons is $ L (L + 1)/2 \mu r^{2}$. In ref.~\cite{Kuti97} $J = L \oplus
S$, $P = \epsilon (-1)^{L+\Lambda +1}$, and $C = \epsilon \eta
(-1)^{L + \Lambda + S}$ with $\epsilon, \eta = \pm 1$. Therefore
with same quantum numbers (L,S), different $J^{PC}$ states are
possible. $L$ and $S$ for these hybrid $J^{PC}$ states are shown
in Table 6 (as given in ref.~\cite{swanson}).
\begin{figure}
\begin{center}
\epsfig{file=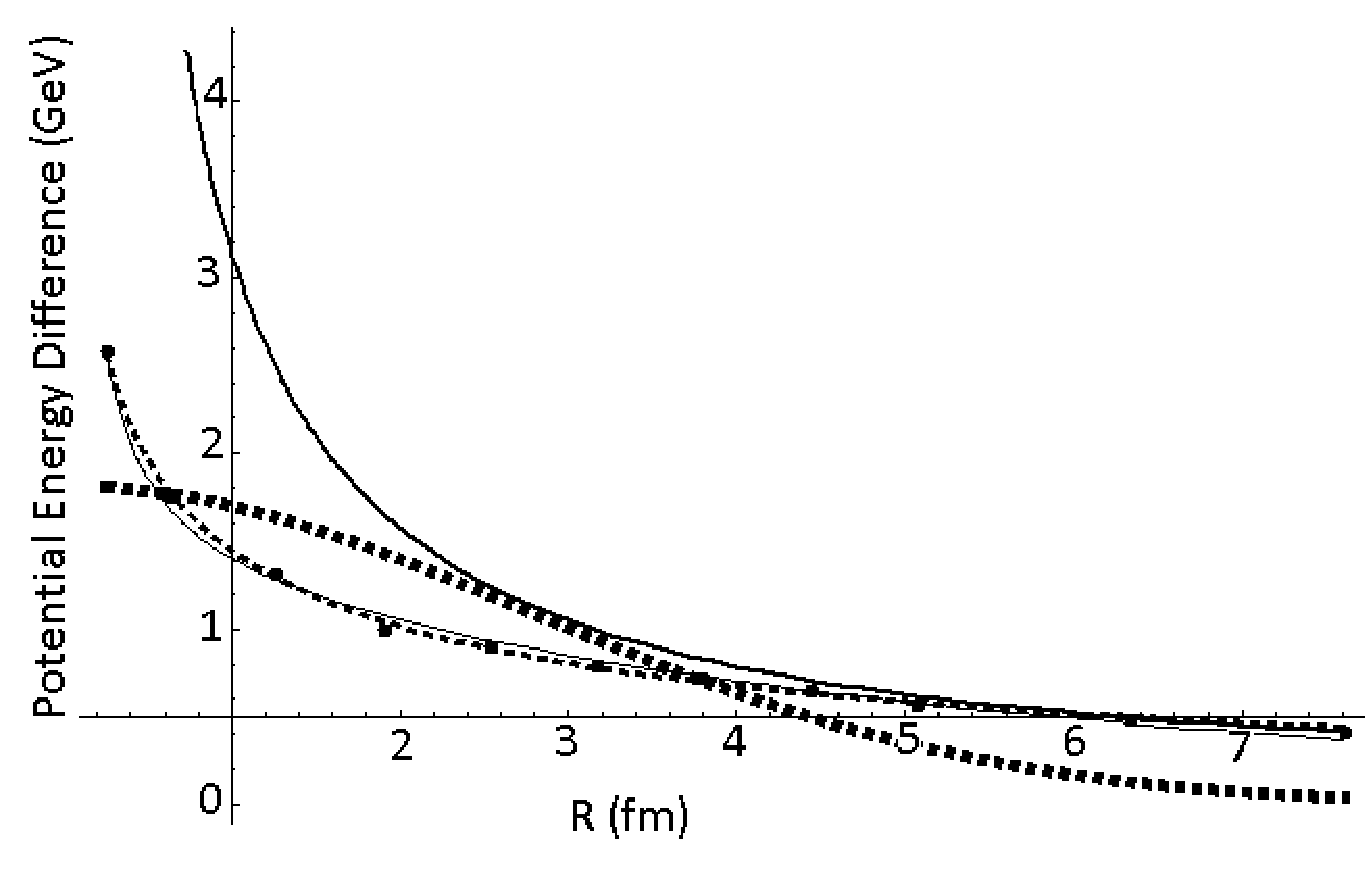,width=0.6\linewidth,clip=}\caption{Graphs of potential energy differences between ground and excited state.
Points represent the data taken from ref.~\cite{Morningstar03}, thick solid line represents potential difference model $\frac{\pi}{r}$, thin solid line represents the potential energy difference $\frac{c}{r} + A \times exp(- B r)$, dashed line represents the potential difference $A \times exp(- B r^{0.1897})$, and squared points represents the potential energy difference $A \times exp(- B r^2)$.}
\end{center}
\end{figure}
For the excited part of quark antiquark potential $\pi/r$ is used in
the flux tube model~\cite{Isgur85}. This form of excitation energy
is only valid at large inter-quark distances. In comparison, we
suggest and evaluate excited potential energy in forms which are valid for smaller distances as well.
For this purpose, we get the potential energy
differences $(\varepsilon_{i})$ between ground and excited states for
different $\emph{r}_{i}$ values from the lattice simulation reported
in Fig. 3 of ref.~\cite{Morningstar03}, and calculate $\chi^{2}$ with a variety
($\pi/r$, $ A \times exp(- B r^{2})$, $ A
\times exp(- B r^{0.1897})$, $\frac{c}{r}+ A \times exp(- B
r^{2})$, and $\frac{c}{r} + A \times exp(- B r^{0.3723})$) of $ans\ddot{a}tz$ by
fitting parameters appearing in each $ans\ddot{a}tz$. Dimensionless $\chi^{2}$ is
defined as
\begin{equation}
\chi^{2} = \frac{\sum^{^{n}}_{_{i=1}}(\varepsilon_{i} - V_{g}(r_i))^{2}}{\sum^{^{n}}_{_{i=1}}\varepsilon_{i}^{2}},\label{P26}
\end{equation}
with $i = 1,2,3,...,n$ being number of data points. Here $V_{g}(r)$
is a model of
the potential energy difference between the ground and excited state. We tried \\
$V_{g}(r) = \pi/r$,  \\
$V_{g}(r) = A \times \textrm{exp}(-B r^2)$, \\
$V_{g}(r) = A \times \textrm{exp}(-B r^{\gamma})$, \\
$V_{g}(r) = \frac{c}{r} + A \times \textrm{exp}(-B r),\hspace{2cm} and$
\\ $V_{g}(r) = \frac{c}{r} + A \times \textrm{exp}(-B r^{\gamma})$.

The parameters $A, B, c,$ and $\gamma$ are found by fitting these
models ($V_g(r)$) with the potential difference data of ref.~\cite{Morningstar03}, and reported in
Table 1. $\chi^2$, $\chi^2/D$ and $\chi$ are also reported in Tables 1 with $\chi^2/D = \chi^2 / (\textrm{Number \hspace{1mm} of \hspace{1mm} data\hspace{1mm} points})$ and $\chi = \frac{\sum^{^{n}}_{_{i=1}} \mid(\varepsilon_{i} - V_{g}(r)) \mid}{\sum^{^{n}}_{_{i=1}} \mid\varepsilon_{i}\mid}$.
\begin{table}\caption{Our calculated
$\chi^{2}$ for the data of potential difference to our suggested models ($V_g (r)$) with best fit parameter's values.}
\begin{center}
\begin{tabular}{|c|c|c|c|c|c|c|c|}
\hline
& & & &\multicolumn{4}{|c|}{Parameters} \\
$ans\ddot{a}tz$ &$\chi^{2}$ & $\chi^2/D$ & $\chi$ &$A$ & $B$ & $c$& $\gamma$ \\ \hline
& & $\textrm{GeV}^2$ & & GeV & GeV &  & \\ \hline
$\pi/r$ & 0.2305 & 0.3268 & 0.5285 & - & - & - & - \\
$A \times \textrm{exp} (- B r^{\gamma})$ & 0.0857 & 0.1215 & 0.2773 & 1.8139 & 0.0657 & - & 2 \\
$A \times \textrm{exp} (- B r^{\gamma})$& 0.0004 & 0.0005 & 0.0205 & 17.9325 & 2.5195 & -& 0.1897 \\
$\frac{c}{r} + A \times \textrm{exp} (- B r^{\gamma})$& 0.0012 & 0.0017 & 0.0331 & 1.2448 & 0.1771& 0.3583  & 1 \\
$\frac{c}{r} + A \times \textrm{exp} (- B r^{\gamma})$& 0.0003 & 0.0004 & 0.0132 & 3.4693 & 1.0110 & 0.1745  & 0.3723 \\ \hline
\end{tabular}
\end{center}
\end{table}

It is noted that
the $\chi^{2}$ for $\pi/r$ is greater than all other potential difference forms. Fig. 2, representing the graphs for different forms of
potential difference, also shows the same behaviour. The $\chi^{2}$
of the gaussian gluonic potential ($ A \times exp(- B r^{2})$) is
less than $\frac{\pi}{r}$, but larger than all other forms. $ A
\times exp(- B r^{2})$ is a smeared form of $\frac{constant}{r}$,
as written in appendix of ref.~\cite{Isgur90}. The potential difference in this
form ($ A \times exp(- B r^{2})$) has an advantage that it can be
easily used in dynamical applications.
 For example, the expectation value
of this part of the potential energy in a Gaussian wave function of the
quadratic confining potential is given in terms of error function
differences even if one uses antiderivatives to evaluate the definite
integrals in it. Or, for usual infinite limits it simply multiplies in the integrand with already Gaussian meson wave functions to keep the integrand as Gaussian, whose well known integral can be written by inspection.
(The
expression for the expectation value can be minimized
with respect to the chosen parameters of the wave function to find
the ground state energy and wave function using the variational
method.)
This calculational advantage may be trivial for a
two-body problem. But if one has to evaluate an expectation value for
few or many body problem (or for a minor variant used in the resonating group
method based treatment~\cite{Nosheen11}~\cite{Imran}
of a system of two quark and two antiquarks), we have to evaluate
an integral of a high order whose direct numerical evaluation may have
convergence problems as in ref.~\cite{Imran}
and a Gaussian integration by inspection may well be the only practical option. The need to keep the multi-dimensional integrals as Gaussian, giving importance to the $ A \times exp(- B r^{2})$ form, becomes even more prominent when when wave functions of the conventional mesons are replaced for the respective problems by those of the hybrids as in ref.~\cite{Nosheen11}.

The analytic Gaussian expectation value of $A \times \textrm{exp}(-B r)$ term is similarly given in terms of error function differences. (Or,  its product in integrand with Gaussian wave functions can be converted to a new Gaussian integrand using a completing of square.)
The $\chi^{2}$ for the potential difference in form of $\frac{c}{r}
+ A \times \textrm{exp}(-B r^{0.3723})$ is much less. The $\frac{c}{r}$ term
in this form can be used for analytical expressions of expectation
values in the above mentioned
Gaussian quark antiquark wave functions of quadratic potential,
resulting in differences of the exponential integral functions
for the most analytical way of finding the expectation values and integrals of resonating group method~\cite{Nosheen11}~\cite{Imran}.
But a similar fully analytical route for Gaussian expectation values and resonating group integrals
of the $A \times \textrm{exp}(-B r^{\gamma})$, for $\gamma$ = non-integer number,
is not available, and this can lead to convergence problems ~\cite{Imran} when we integrate numerically integrals of high dimensions.
\begin{table}\caption{Our calculated
$\chi^{2}$ for the first excited state data of ref.~\cite{Morningstar03} to the model used in ref.~\cite{Morningstar97} with best fit parameter's values.}
\begin{center}
\begin{tabular}{|c|c|c|c|c|c|c|c|}
\hline
& & & &\multicolumn{4}{|c|}{Parameters} \\
$\textrm{Excited} \hspace{1mm} \textrm{Potential}$
&$\chi^{2}$ & $\chi^2/D$ & $\chi$ &$b_0$ & $b_1$ & $b_2$& $c_0$ \\ \hline
& &$\textrm{GeV}^2$ & &$\textrm{GeV}^2$& $\textrm{GeV}^3$ & $\textrm{GeV}^4$ & GeV \\ \hline
$c_0 + $ & 0.00096 & 0.0030 & 0.0226 & 58.0016 & $4.4896 \times $ & 0.2859 & -6.1814 \\
$\sqrt{b_0 + b_1 r + b_2 r^2}$ & & &  &  & $10^{-10}$ &  & \\ \hline
\end{tabular}
\end{center}
\end{table}
\begin{figure}
\begin{center}
\epsfig{file=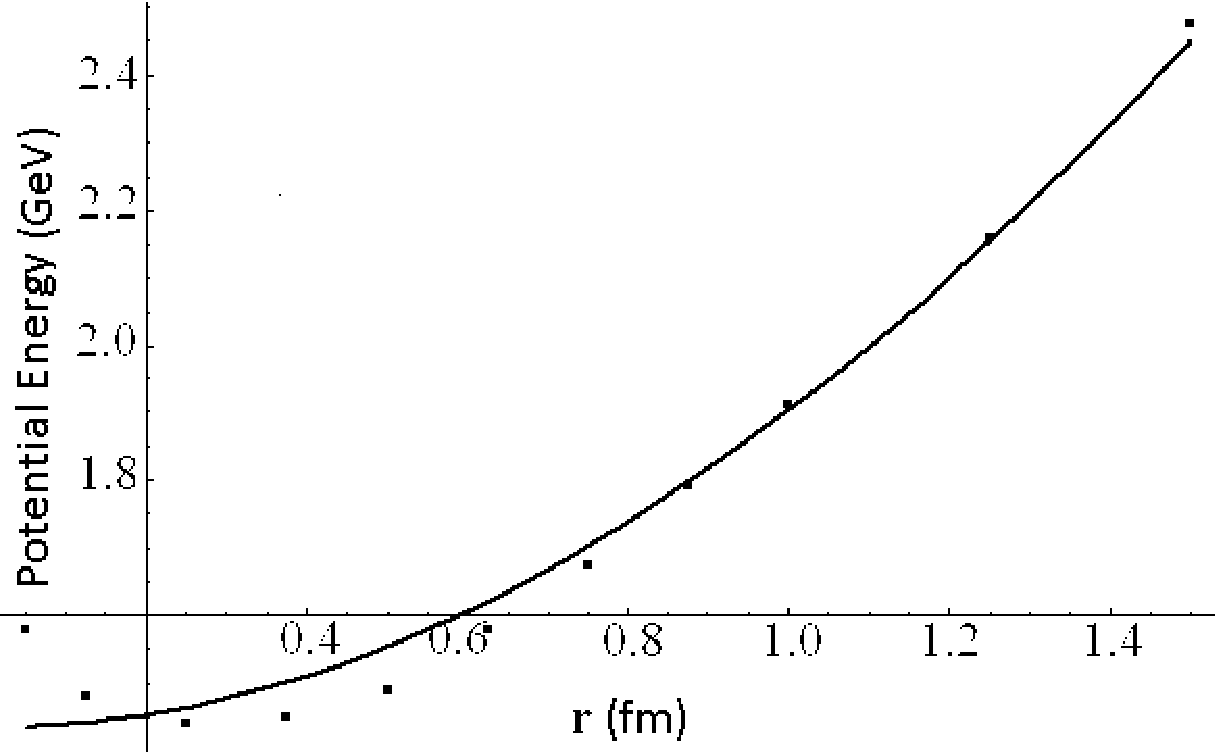,width=0.6\linewidth,clip=}\caption{Graphs of excited state potential energy in the form of $c_0 + \sqrt{b_0 + b_1 r + b_2 r^2}$ along with data of ref.~\cite{Morningstar03}. Solid line is for the potential $c_0 + \sqrt{b_0 + b_1 r + b_2 r^2}$ and dots are for data of ref.~\cite{Morningstar03}.}
\end{center}
\end{figure}
\begin{figure}
\begin{center}
\epsfig{file=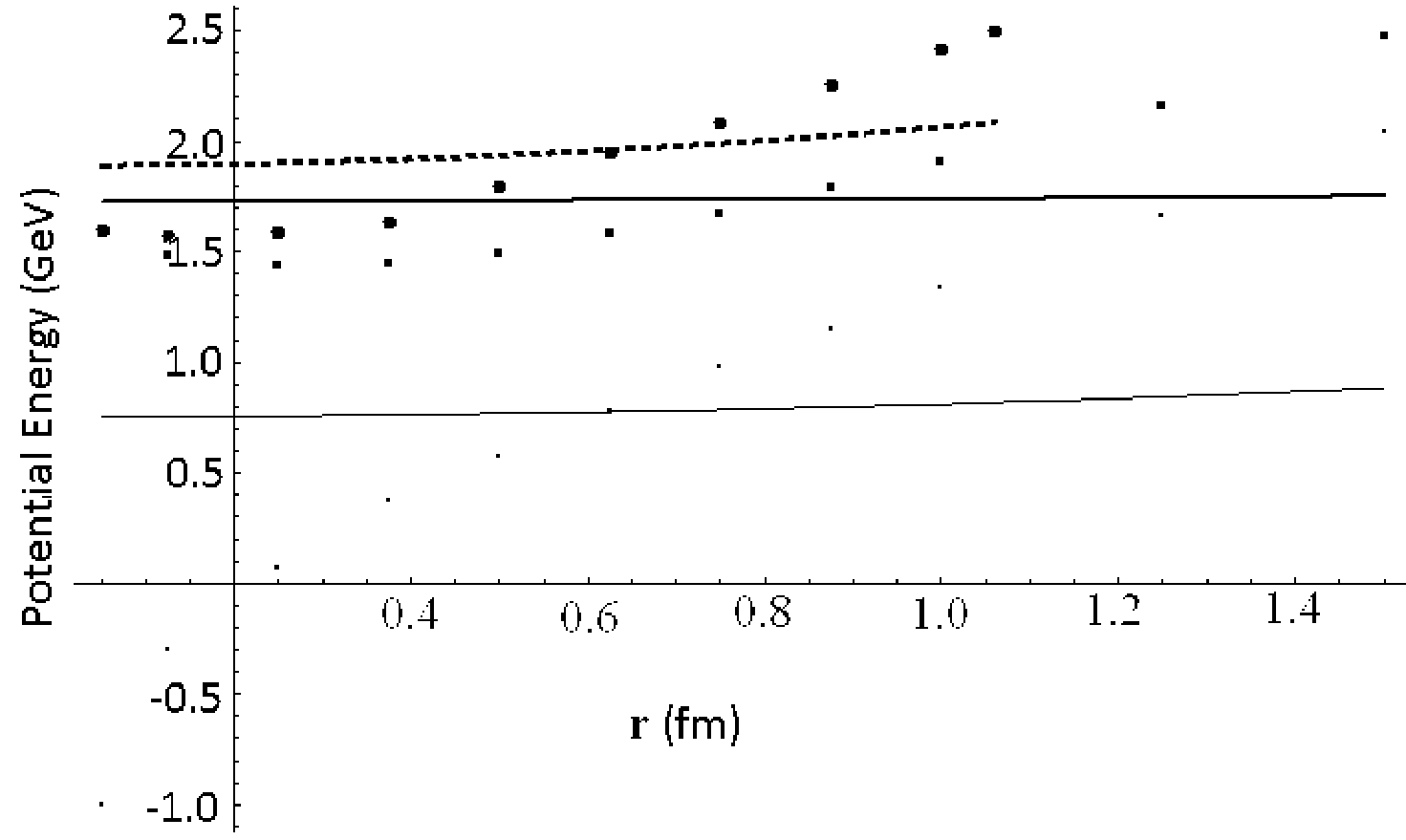,width=0.6\linewidth,clip=}\caption{Graphs of potential energy in the form of $\sqrt{\sigma^2 r^2 + 2 \pi \sigma (N + \frac{3}{2}) + w_q}$.
The solid line is for $N = 0$ and small points are for $N = 0$ data of ref.~\cite{Morningstar03}. Thick solid line is for $N = 1$ and medium size points are for $N = 1$ data of ref.~\cite{Morningstar03}. Dashed line is for the $N = 2$ and large size points are for $N = 2$ data of ref.~\cite{Morningstar03}.}
\end{center}
\end{figure}

In ref.~\cite{Morningstar97},~\cite{Kuti97}
 excited state potential ($\textit{not}$
 the difference) is used in the form of
 \begin{equation}
c_0 + \sqrt{b_0 + b_1 r + b_2 r^2}.
\end{equation}
By fitting this potential with data of first excited potential taken from Fig.3 of ref.~\cite{Morningstar03}, we calculated the parameters $b_0$, $b_1$, $b_2$, $c_0$. We also calculated
$\chi^2$, $\chi$ and $\chi^2/D$
and reported in Table 2 along with the parameter values.
The fit of the data with this excited state potential is shown in our Fig.3. As for the the analytical calculations (for finding expectation values etc.), these are also not possible with this potential form and thus for many applications it has to be replaced by others of higher chisquare.

Fig.4 shows the behaviour of string
potential~\cite{Buisser}
\begin{equation}
\sqrt{\sigma^2 r^2 + 2
\pi \sigma (N + \frac{3}{2}) + w_q}
\end{equation}
with $N = 0,1,2$. In this figure, the points represent the data
taken
from ref.~\cite{Morningstar03}.
The parameter's values, $\chi^2$ and other error measures
of the string potential
with the data (of excited potential)
of ref.~\cite{Morningstar03} is reported in Table 3. The parameter's values are calculated by fitting the data of excited potential with the string potential. The
analytical calculations (for finding expectation values) are not possible with this string potential form as well.
\begin{table}\caption{Our calculated
$\chi^{2}$ with best fit parameter's values.}
\begin{center}
\begin{tabular}{|c|c|c|c|c|c|c|c|c|c|}
\hline
& & & &\multicolumn{3}{|c|}{Parameters} \\
$Excited \hspace{1mm} Potential$ &$\chi^{2}$ & $\chi^2/D$ & $\chi$ &$N$ & $w_q$ & $\sigma$\\ \hline
& &$\textrm{GeV}^2$ & GeV& - & $\textrm{GeV}^2$ &$\textrm{GeV}^4$\\ \hline
 &0.5536 &0.6633 & 0.6932 & 0 & $5.9884 \times 10^{-9}$ & 0.0601 \\
$\sqrt{\sigma^2 r^2 + 2
\pi \sigma (N + \frac{3}{2}) + w_q}$ &0.0112 &0.0348 & 0.0899 & 1 & $3.5044 \times 10^{-7}$ & 0.1629 \\
 &0.0190 &0.0740 & 0.1259 & 2 & $1.0531 \times 10^{-5}$ & 0.1627 \\ \hline
\end{tabular}
\end{center}
\end{table}

\subsection*{Wave Functions and Radii of Hybrid Charmonium Mesons} \qquad
Now, we can write the quark antiquark
potential in excited state gluonic field as
\begin{equation}
V(r)= \frac{-4 \alpha_{s}}{3 r}+ b\, r + \frac{32 \pi \alpha_{s}}{9 m^{2}_{c}}
(\frac{\sigma}{\sqrt{\pi}})^{3} e^{-\sigma^{2} r^{2}} S_{c}.S_{\overline{c}}+ \frac{L (L + 1)
- 2\Lambda^{2} + <J^{2}_{g}>}{2 \mu r^{2}} + V_{g}(r).\label{P27}
\end{equation}
$V_{g} (r)$ is defined above after eq.\eqref{P26}.

Using this excited state potential of eq.\eqref{P27} along with the
above mentioned values (after eq.\eqref{P28}) of $\Lambda$ and
$<J^{2}_{g}>$, the energy eigenvalues and the corresponding wave
functions are found by using the same technique as employed for
conventional mesons (mentioned in section II). As before, these eigenvalues
plus constituent quark antiquark masses are taken to be the masses
of hybrid mesons. Then we normalized the wave functions and found the
root mean square radii of hybrid mesons by using eq.\eqref{P25}. The
normalized radial wave functions
for charmonium hybrid mesons are graphically
represented in Fig.5 and Fig.6.
 The overlaps
  of our numerically calculated radial
wave functions ($U = r R$) for the excited states and a modified
 gaussian wave function ans$\ddot{\textrm{a}}$tz
 \begin{equation}
 \psi =n\, r^2 \textrm{exp}(- p r^2) \label{exf}
 \end{equation}
multiplied by $\sqrt{4 \pi} r$ are written in Table 4  in such a way that $U = r R = \sqrt{4 \pi} r \psi$.

The normalization of the gaussian wave function gives
\begin{align}
n = (4 2^{\frac{3}{4}}p^{\frac{7}{4}})(15^{\frac{1}{2}}\pi^{\frac{3}{4}}).\label{np}
\end{align}
The numerical value of $p$ is found by fitting this function with the data of numerically calculated wave function, and written in
Table 4.
\begin{table}\caption{Our best fit parameter's values and overlaps of numerically solved wave functions
 to the wave function form written in eq.\eqref{exf}.}
\begin{center}
\begin{tabular}{|c|c|c|c|c|c|c|c|}
\hline
& \multicolumn{3}{|c|}{$L=1$ and $S=1$} & \multicolumn{3}{|c|}{$L=2$ and $S=1$}  \\
& \multicolumn{2}{|c|}{Parameters of excited state} & Overlap & \multicolumn{2}{|c|}{Parameters of excited state} & Overlap\\
& \multicolumn{2}{|c|}{function written in eq.\eqref{exf}} & & \multicolumn{2}{|c|}{function written in eq.\eqref{exf}} &\\
$ans\ddot{a}tz$ & $n$ & $p$ & & $n$ & $p$ &\\ \hline
$\pi/r$ & 0.0047 & 0.0556 & 0.9997 & 0.0032 & 0.0451 & 0.9989 \\
$A \times \textrm{exp} (- B r^{2})$ & 0.0031 & 0.0439 & 0.9964 & 0.0025 & 0.0391 & 0.9905 \\
$A \times \textrm{exp} (- B r^{0.1897})$ & 0.0093 & 0.0822 & 0.9903 & 0.0048 & 0.0565 & 0.9998 \\
$\frac{c}{r} + A \times \textrm{exp} (- B r)$ &  0.0070 & 0.0698 & 0.9899 & 0.0036 & 0.0482 & 0.9999\\
$\frac{c}{r} + A \times \textrm{exp} (- B r^{0.3723})$ & 0.0087 & 0.0794 & 0.9903 & 0.0045 & 0.0547 & 0.9998 \\
$c_0 + \sqrt{b_0 + b_1 r + b_2 r^2}$& 0.0163& 0.1135 & 0.9921 & 0.0090 & 0.0807 & 0.9999 \\ \hline
\end{tabular}
\end{center}
\end{table}

\begin{figure}
\begin{center}
\epsfig{file=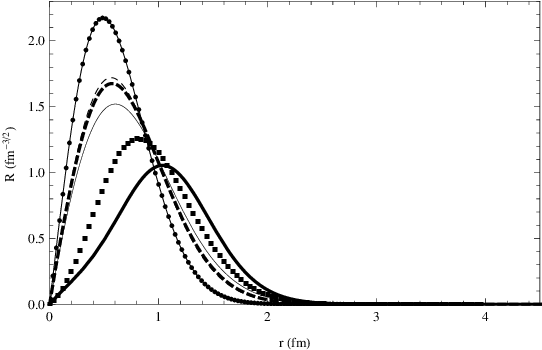,width=0.9\linewidth,clip=}\caption{Hybrid charmonium meson radial wave functions for $0^{+ -}, 1^{- +}$ and $2^{+ -}$ $J^{PC}$ states for L=1 and S=1. The wave function with potential in the form of coulombic plus linear plus $Aexp(- B r^2)$ is represented by the solid line. Wave function with coulombic plus linear plus $\pi/r$ potential is represented by square points. Wave function with coulombic plus linear plus $A \times Exp(-B\, r^{0.1896})$ potential is represented by dashed line. Wave function with coulombic plus linear plus $\frac{c}{r} + A \times Exp(-B\, r)$ potential is represented by thin solid line. Wave function with coulombic plus linear plus $\frac{c}{r} + A \times Exp(-B\, r^{0.3723)}$ potential is represented by thick dashed line, and the wave function with excited potential in the form of $c_0 + \sqrt{b_0 + b_1 r +b_2 r^2}$ is represented by points with lines.}
\end{center}
\end{figure}
\begin{figure}
\begin{center}
\epsfig{file=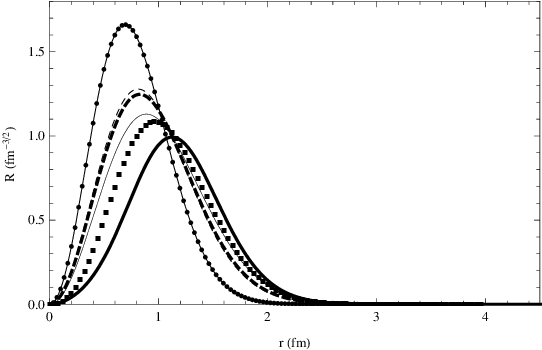,width=0.9\linewidth,clip=}\caption{The same figure as Fig.4 but with L = 2 and S = 1.}
\end{center}
\end{figure}
The Fig.5 and Fig.6 show the wave function dependence on $L$ and
$S$. Therefore the masses and root mean square radii of $0^{+ -}, 1^{-
+}$ and $2^{+ -}$ $J^{PC}$ states also depend on the quantum numbers
$L$ and $S$. $|R(0)|^2$ is found for hybrid mesons using
eq.\eqref{mid}. The Fig.5 and Fig.6 also show that the peaks of the graphs for radial probability density are in agreement with the peak
 of radial probability curve drawn in ref.~\cite{Morningstar03}.

\section*{IV. Results and Conclusions}

\qquad 1. For conventional mesons,
 our calculated masses and root mean square radii are
reported in Table 5 along with the experimental and theoretical
predictions of the other works. We observed that our results are in
good agreement with the experimental and existing theoretically
predicted values, which shows the validity of our method. Quantum
mechanically, when L increases, centrifugal  barrier increases so
particles become less bound implying increased root mean square
radii. Our calculated root mean square radii are in agreement with
this expectation.

2. With the parameters (given in Table 6) for the $ 0^{+ -}, 1^{-
+}$ and $2^{+ -}$ $J^{PC}$ states, masses and root mean square radii
are calculated for the charmonium hybrid mesons. In Table 6, masses
are calculated using the excited state gluonic field potential in
the above mentioned forms. For comparison with earlier works, masses
of $c\overline{c}$ hybrid mesons with $ 0^{+ -}, 1^{- +}$ and $2^{+
-}$ $J^{PC}$ states are given in Table 7. In Table 8, root mean
square radii are calculated by taking the excited state potential in
the coulomb plus linear plus additional excited potential. In Table 9, masses and root mean square radii are reported for the excited potential in the form of $c_0+ \sqrt{b_0 + b_1 r + b_2 r^2}$.
\begin{table}\caption{The experimental and theoretical masses and theoretical root mean square radii of some conventional
charmonium mesons.
The experimental mass is the average PDG~\cite{charmonia05} and rounded to 0.001 GeV. Our calculated masses are rounded to 0.0001 GeV.}
\begin{center}
\begin{tabular}{|c|c|c|c|c|c|c|c|c|}
\hline
Meson & $L$ & $S$ & Our calculated & Theor. mass~\cite{charmonia05} & Exp. mass & our calculated & Theor. $\sqrt{\langle r^{2} \rangle}$~\cite{size} \\
& & & mass & with NR potential & & $\sqrt{ \langle r^{2} \rangle}$ & with \\
& & & & model & &  & potential model \\ \hline
& & & \textrm{GeV} & \textrm{GeV} & \textrm{GeV} & \textrm{fm} & \textrm{fm}\\ \hline
$ \eta_{c}$ & 0 & 0 & 2.9816 & 2.982 &$2.9792\pm 0.0013~\cite{charmonia05}$ & 0.365 & 0.388\\ \hline
$J/\psi$& $0$ & $1$ & $3.0900$ & $3.090$ & $3.09687\pm 0.00004$~\cite{charmonia05}&$0.414$ & $0.404$ \\ \hline
$h_{c} $&  1 & $0$ &  $3.5156$ & $3.516$ & $3.525\pm 0.00055$~\cite{charmonia05}  & $0.674$ & $0.602$\\  \hline
$\chi_{c}$ & $1$ & $1$ & $3.5246$& $3.556$ & $3.55618\pm 0.00013$~\cite{Seth05}& $0.685$ & $0.606$\\
\hline
\end{tabular}
\end{center}
\end{table}
\begin{table}\caption{Our calculated
masses of $c\overline{c}$ hybrid meson $ 0^{+ -}, 1^{- +}$ and $2^{+
-}$ $J^{PC}$
 states.}
\begin{center}
\begin{tabular}{|c|c|c|c|c|c|c|c|c|c|c|}
\hline
$J^{PC}$&$L$&$S$&$\Lambda$&$<J^{2}_{g}>$ &\multicolumn{5}{|c|}{Excited potential as coulombic plus linear plus} \\ \hline
& & & & & $\pi/r$ & $A \times $ & $A \times $ & $\frac{c}{r}+ A \times $ & $\frac{c}{r} + A \times $ \\
& & & & & & $\textrm{exp} (- B r^2)$ & $\textrm{exp} (- B r^{0.1897})$ & $\textrm{exp} (- B r)$ & $\textrm{exp} (- B r^{0.3723})$ \\ \hline
& & & & & $ \textrm{GeV}$ & $\textrm{GeV}$ & $\textrm{GeV}$ &
$\textrm{GeV}$ &$\textrm{GeV}$ \\ \hline
$ 0^{+ -}, 1^{- +}, 2^{+ -} $ & 1 & 1 & 1 & 2 & 4.3571 & 4.0619 & 4.2680 & 4.2733 & 4.2694 \\
$1^{- +}, 2^{+ -}$ & 2 & 1 & 1 & 2 & 4.4632 & 4.1433 & 4.4632 & 4.4258 & 4.40796 \\
\hline
\end{tabular}
\end{center}
\end{table}
\begin{table}\caption{The mass predictions of $1^{- +}, 0^{+ -}$ and $2^{+ -}$ states of other works.}
\begin{center}
\begin{tabular}{|c|c|c|c|}
\hline
\multicolumn{3}{|c|}{Predicted masses (\textrm{GeV})} & models\\ \hline
$ 1^{- +}$ & $0^{+ -}$ & $2^{+ -}$ & \\ \hline
$ \approx 3.9$~\cite{Kuti80} & & & bag model\\ \hline
4.2-4.5~\cite{Paton85}-\cite{Paton852} & & & flux tube model \\ \hline
$4.19 \pm sys. error ~\cite{Griffiths} ~\cite{Michael90}$& $\approx 4.5$~\cite{Morningstar97} & $\approx 4$~\cite{Morningstar97} & heavy quark LGT \\
$4.7 ~\cite{Iddir}$&4.58~\cite{Iddir} &  &  \\  \hline
4.1-4.5 & & & QCD sum rules \\ \hline
$4.369-4.420~\cite{Mankeetal,Mei03,Liu06}$ & 4.714(52)~\cite{Liu06} & 4.895(88)~\cite{Liu02}& quenched lattice QCD\\
\hline
\end{tabular}
\end{center}
\end{table}
\begin{table}\caption{Our calculated
root mean square radii of $c\overline{c}$ hybrid meson $ 0^{+ -},
1^{- +}$ and $2^{+ -}$ $J^{PC}$
 states.}
\begin{center}
\begin{tabular}{|c|c|c|c|c|c|c|c|c|c|c|}
\hline
$J^{PC}$&$L$&$S$&$\Lambda$&$<J^{2}_{g}>$ & \multicolumn{5}{|c|}{$\sqrt{\langle r^{2} \rangle}$ with excited potential as coulombic plus linear plus} \\ \hline
& & & & & $\pi/r$ & $A \times $ & $A \times $ & $\frac{c}{r}+ A \times $ & $\frac{c}{r} + A \times $ \\
& & & & & & $\textrm{exp} (- B r^2)$ & $\textrm{exp} (- B r^{0.1897})$ & $\textrm{exp} (- B r)$ & $\textrm{exp} (- B r^{0.3723})$ \\ \hline
& & & & & $\textrm{fm}$ & $\textrm{fm}$ & $\textrm{fm}$ & $\textrm{fm}$ &$\textrm{fm}$\\ \hline
$ 0^{+ -}, 1^{- +}, 2^{+ -} $ & 1 & 1 & 1 & 2 & 1.1061 & 1.2458 & 0.9110 &0.9881 & 0.9272 \\
$1^{- +}, 2^{+ -}$ & 2 & 1 & 1 & 2 & 1.2280 & 1.3203 & 1.0988& 1.1883 & 1.1160 \\
\hline
\end{tabular}
\end{center}
\end{table}

\begin{table}\caption{Our calculated masses and root mean square radii with excited potential $ c_0 + \sqrt{b_0 + b_1 r + b_2 r^2}$ of $c\overline{c}$ hybrid meson with $ 0^{+ -}, 1^{- +}$, $2^{+ -}$ $J^{PC}$ states.}
\begin{center}
\begin{tabular}{|c|c|c|c|c|c|c|}
\hline
$J^{PC}$&$L$&$S$&$\Lambda$&$<J^{2}_{g}>$ & masses &  $\sqrt{\langle r^{2} \rangle}$\\
& & & & & $\textrm{GeV}$ & $\textrm{fm}$ \\ \hline
$ 0^{+ -}, 1^{- +}, 2^{+ -} $ & 1 & 1 & 1 & 2 & 4.9503 & 0.7748 \\
$1^{- +}, 2^{+ -}$ & 2 & 1 & 1 & 2 & 5.1693 & 0.9185 \\
\hline
\end{tabular}
\end{center}
\end{table}

3. For conventional mesons
 $|R(0)|^{2}$ is reported in Table 10.
Each $|R(0)|^{2}$ of $c\overline{c}$ hybrid mesons for $ 0^{+ -},
1^{- +}$ and $2^{+ -}$ $J^{PC}$ states is equal to zero by our
calculation and this result agrees with ref.~\cite{Stephen} which
writes "models of hybrids typically expect the wave function at the
origin to vanish". We also noted that the masses and root mean
square radii of the hybrid mesons are greater than ordinary mesons
with the same flavour and quantum numbers ($L$ and $S$).
Since $0^{+-},1^{-+},\hspace{1mm}2^{+-}$ states are not possible with
quark model quantum numbers, so we can not compare these $J^{PC}$
states with conventional mesons.

In ref.~\cite{Ananthanarayan}, the scalar form factor is written
as
\begin{equation}
\overline{\Gamma}_{\pi}(t)=1 + \frac{1}{6}\langle r^2 \rangle_s^{\pi} t + o(t^2). \label{scalar}
\end{equation}
In ref.~\cite{S. I. Kruglov}, energy shift and magnetic
polarizability are written
as
\begin{equation}
\bigtriangleup E_n= \langle n \mid \frac{e H}{8 \mu} L_3 + \bigg( \frac{e^2}{4 \widetilde{\mu}} +
\frac{q^2}{\mu_1 + \mu_2}\bigg)\frac{H^2 r^2 sin^2 \theta}{32}\mid n' \rangle +
\Sigma_n'\frac{\mid \langle n'\mid e H L_3/8 \widetilde{\mu} \mid n \rangle \mid ^2}{E_n - E_{n'}}. \label{energy}
\end{equation}
Here the symbol $L_3$,$H$, $e$, $m$ are used for the angular momentum, magnetic field, charge and mass of the quark, $\widetilde{\mu} = \frac{\mu}{2}$, $\theta$ is the angle between $H$ and relative co-ordinate
$r$, $q = e_1 -e_2$, and $e = e_1 +e_2$.
In above equation, the term having $\langle n \mid \frac{H^2 r^2 sin^2 \theta}{32} \mid n' \rangle$ is related to square of root mean square radii.
\begin{equation}
\beta = -\frac{1}{24}\bigg( \frac{e^2}{4 \widetilde{\mu}} + \frac{q^2}{\mu_1 + \mu_2}\bigg)\langle r^2 \rangle. \label{magnetic}
\end{equation}
In above eqs.(\ref{scalar}-\ref{magnetic}) the
root mean square radii is in the numerator,
therefore we predict that magnitudes of scalar form
factor~\cite{Ananthanarayan}, energy shift~\cite{S. I. Kruglov}, and magnetic
polarizability ~\cite{S. I. Kruglov} for hybrids are greater than those for
conventional mesons of the same quantum numbers ($L$ and $S$).

By parametrizing the excited state wave function written above in eq.\eqref{exf},
we get
\begin{equation}
\langle r^2 \rangle = \int \psi^{\ast} r^2 \psi\, d r = \frac{15 n^2 \sqrt{\pi/2}}{128 p^{7/2}}.\label{nrp}
\end{equation}
Here $n$ is a function of $p$, as given by eq.\eqref{np}.
Substituting the result of eq.\eqref{nrp} in
eqs.(\ref{scalar},\ref{magnetic}), the scalar form factor and
magnetic polarizability become
\begin{equation}
\overline{\Gamma}_{\pi}(t)=1 + \frac{1}{6} \frac{15 n^2 \sqrt{\pi/2}}{128 p^{7/2}} t + o(t^2), \label{scalar1}
\end{equation}
\begin{equation}
\beta = -\frac{1}{24}\bigg( \frac{e^2}{4 \widetilde{\mu}} + \frac{q^2}{\mu_1 + \mu_2}\bigg)\frac{15 n^2 \sqrt{\pi/2}}{128 p^{7/2}}.\label{mag}
\end{equation}
Numerically calculated values of $n$ and $p$ are written in Table 4
for different forms of excited state potentials for different $L$ and $S$.

As we mentioned above, $|R(0)|^{2}$ is equal to zero for hybrid
mesons. Using this result, we can predict that decay
constants~\cite{Bhavin}, decay rates~\cite{Bhavin}, and differential
cross sections~\cite{diff} of hybrid mesons are zero as these
quantities are proportional to $|R(0)|^{2}$ as written in these references.

\begin{table}\caption{$|R(0)|^{2}$ of $c\overline{c}$ meson}
\begin{center}
\begin{tabular}{|c|c|c|c|}
\hline
Meson & $L$ & $S$ & our calculated normalized $|R(0)|^{2}$\\
& & & $GeV^{3}$\\ \hline
$ \eta$ & 0 & 0 &  1.2294 \\ \hline
$J/\psi$& $0$ & $1$ & 1.9767\\ \hline
$h_{c} $&  1 & $0$ & $ \approx 0$\\  \hline
$\chi_{c}$ & $1$ & $1$ &  $ \approx 0$\\
\hline
\end{tabular}
\end{center}
\end{table}

\section*{Acknowledgement}
\qquad We are grateful to Higher education Commission of Pakistan
for their financial support no.17-5-3 (Ps3-212) HEC/Sch/2006.

\end{document}